\documentclass[twocolumn,12pt]{iopart}
\usepackage{iopams,graphicx}
  
\begin{document}

\title {CPA for strongly correlated systems: Electronic structure and magnetic properties of NiO-ZnO solid solutions.}

\author {M A Korotin$^1$, Z V Pchelkina$^{1,2}$, N A Skorikov$^1$, E Z Kurmaev$^1$ and V I Anisimov$^{1,2}$}

\address {$^1$Institute of Metal Physics, 620990, Ekaterinburg, Russia}
\address {$^2$Ural Federal University, 620002, Ekaterinburg, Russia}
\ead{pchelkzl@mail.ru}

\begin {abstract}

The method of electronic structure calculations for strongly correlated disordered materials 
is developed employing the basic idea of coherent potential approximation (CPA). Evolution of electronic structure 
and spin magnetic moment value with concentration $x$ in strongly correlated Ni$_{1-x}$Zn$_x$O solid solutions is investigated 
in the frame of this method. The obtained values of energy gap and magnetic moment are in agreement with the 
available experimental data.

\end {abstract}


\maketitle

\section {Introduction}

The study of the electronic structure of the solid compounds with the impurities within 
density functional theory could be performed in two ways. One is the supercell method when 
the unit cell of parent compound is enlarged by several times and one of the atoms is 
replaced by the atom of impurity. This method has two basic shortcomings: 
first, the atoms of impurity are turned out to be ordered and second, only discrete concentration could 
be treated (specified by the number of unit cells). Another approach is the coherent potential approximation developed 
in the 1970s~\cite{cpa}. The main idea of the CPA is the modification of the electronic Green function by introducing 
the coherent potential describing self-consistent effective medium containing randomly distributed impurity. 
The compounds with arbitrary concentration and disordered impurities could be naturally treated within this approach. 
The main objects of investigation studied within CPA until now are the transition metal and semi-conductor alloys (for instance Ga-As), 
which generally have quite similar potentials of matrix and impurity atoms. Recently the CPA was also applied for 
nonstoichiometric materials with vacancies in metallic and nonmetallic sublattices (TiO$_{2-\delta}$, TiO$_y$, 
TiO$_{2-x-y}$C$_x$N$_y$)~\cite{ourcpa, tio_cpa, tiocn_cpa, tiocn_cpa2}. 

In searching the alloy theory based on first principles the large efforts have been made to combine CPA with 
different methods of electronic structure calculation in the 1980s and 1990s. The Korringa-Kohn-Rostoker (KKR) method was combined with 
CPA approach and applied to the Cu-Ni alloy~\cite{Faulkner}. The KKR-CPA was found to be very time consuming. The well known 
problems with the energy dependence of the structure constants which takes most of calculation time can be solved by linear approaches 
in the band theory of ordered systems like linearized muffin-tin orbitals method (LMTO). 
The next step was the realization of TB-LMTO-CPA method proposed by Kudrnovsk\'{y} and Drchal~\cite{Kudrnovsky} though some 
difficulties with the Brillouin zone integration remain. The efficient scheme for obtaining the electronic properties of 
random solid solutions was developed on the basis of CPA and LMTO Green function method by Abrikosov et al. 
and applied for Cu-Ni and Cu-Au alloys~\cite{Abrikosov}. 
In the last few years the interest to the CPA approach is revived mainly because of its natural incorporation into the LDA+DMFT 
computational scheme. Such combination gives opportunity for investigation of strongly correlated disordered systems. 
The topical review of the fully self-consistent KKR-based Local Spin Density Approximation + Dynamical Mean Field Theory 
(LSDA+DMFT) scheme and its application to the ordered and disordered compounds 
was performed by Min\'{a}r~\cite{Minar11}. This scheme was successfully employed to Fe$_x$Ni$_{1-x}$ alloy~\cite{Minar05}, 
Co-Pt solid-state systems~\cite{Sipr08} and magnetic alloy system Ni$_x$Pd$_{1-x}$~\cite{Braun10}. 

In the present paper we develop the scheme for calculation of electronic structure of strongly correlated disordered 
systems based on the CPA. This method is applied for investigation of electronic structure and magnetic moment of the strongly correlated NiO-ZnO solid solutions. 
The comparison with available experimental data is performed. The current scheme is conceptually close to the dynamical mean field theory DMFT~\cite {DMFT}. 
In present version the strong correlations are treated in the static mean-field limit, while the fully 
self-consistent CPA-DMFT method is the challenge for the future work. 

\section {CPA method}

The comprehensive review on the CPA methodology could be found in \cite{CPA_rev}. Below the basic equations of current 
CPA scheme are presented.
The self-consistent coherent potential approximation cycle starts from the calculation 
of single-site electronic Green function $\hat{G}(i\omega_n)$ on the imaginary axis:
\begin{equation}
\hat G(i\omega_n) = \sum_k \big( (\mu + i\omega_n) \hat {\mathbf 1} - \hat H_0(k) -\hat \Sigma (i\omega_n) \big) ^{-1},
\label{eq:G_iw}
\end{equation}
where $\omega_n = (2n+1) \pi /\beta, n=0, \pm 1, \pm 2$,\ldots\ are the odd Matsubara frequencies, 
$\beta$=1/T is inverse temperature, $\mu$ is a chemical potential, $\hat {\mathbf 1}$ is the unit matrix, 
$\hat H_0(k)$ is the Hamiltonian of the system without impurity, 
the summation runs over {\bf k} vectors of reciprocal lattice in the irreducible part of the Brillouin zone, 
$\hat \Sigma(i\omega_n)$ is starting coherent potential. 
As in the dynamical mean-field theory $\hat \Sigma$ is assumed to be {\bf k} independent which is correct in 
the limit of $d \rightarrow \infty$ and hence all calculated Green functions are wave vector independent.

The self-consistent solution is achieved if the single-site Green function (\ref{eq:G_iw}) coincide with the 
Green function of single impurity model $\hat G = \hat G_{imp}$. 
Let us call the part of the Green function corresponding to effective medium (the sites on which the 
impurity is uniformly distributed) as $\hat G_{eff}$. Then the Dyson equation for the Green function of the 
bath will have the form
\begin{equation}
\hat G_0 = \big( \hat G_{eff}^{-1}+\hat \Sigma \big) ^{-1}.
\label{eq:G_0}
\end{equation}
The effective Green function at the sites containing the impurity with concentration $x$ has the form
\begin{equation}
\hat G_{eff} = (1-x) \hat G_{host} + x \hat G_{imp},
\label{eq:G_new}
\end{equation}
where 
\begin{equation}
\hat G_{host} = \hat G_0,
\label{eq:G_host}
\end{equation}
\begin{equation}
\hat G_{imp} = \hat G_0 \big( 1-\Delta V \hat G_0 \big) ^{-1},
\label{eq:G_imp}
\end{equation}
and $\Delta V$ is the difference in potentials between the ``impurity'' and ``host'' atoms.
The equation (\ref{eq:G_new}) replaces the solution of the single impurity Anderson model 
in DMFT method. New coherent potential could be obtained from equation (\ref{eq:G_0}):
\begin{equation}
\hat \Sigma^{new} = - \big( \hat G_{eff} \big) ^{-1} + \big( \hat
G_0 \big) ^{-1}.
\label{eq:Sigma_new}
\end{equation}
Then the new coherent potential is used in equation (\ref{eq:G_iw}) and procedure of 
searching of $\hat \Sigma$ is repeated until the self-consistency condition will be fulfilled. 
Finally, to calculate the density of states (DOS), the
self-consistent coherent potential $\hat \Sigma(i\omega)$ is analytically
continued to the real energy axis $\epsilon$ using the Pad\'e approximants 
\cite {Pade}.

This scheme was developed and successfully applied to the investigation of the electronic
structure of nonstoichiometric rutile~\cite{ourcpa}, titanium monoxide~\cite{tio_cpa}, 
and titanium dioxide with carbon and nitrogen impurities in oxygen
sublattice \cite {tiocn_cpa, tiocn_cpa2}. In the case of titanium dioxide 
the scissors operator was employed to adjust the calculated energy gap value to the experimental one. 

The new feature of the current version of CPA is the treatment of the strongly correlations 
in the static mean-field limit~\cite{LDA+U}. This scheme is appropriate for disordered compounds with 
strongly correlated host sublattice (like NiO). In this case   
equation (\ref {eq:G_host}) is transformed in the same manner as (\ref{eq:G_imp}) to
\begin {equation}
\hat G_{host}=\hat G_0 (1-\Delta V^{Coulomb} \hat G_0)^{-1}
\label {eq:G_host_U}
\end {equation}
where $\Delta V^{Coulomb}$ could be found in the static mean-field limit~\cite{LDA+U} as
\begin {eqnarray}
%\begin {split}
&\Delta V^{Coulomb}=V_{mm'}^\sigma =  \nonumber\\
&\sum_{m''m'''} \{\langle m,m'' |V_{ee}| m',m'''\rangle n_{m''m'''}^{-\sigma } +  \nonumber\\
&[\langle m,m''|V_{ee}| m',m''' \rangle - \langle m,m'' |V_{ee}| m''',m' \rangle] n_{m'' m'''}^\sigma \} - \nonumber\\ 
&U (n-\frac 12) + J (n^{\sigma}-\frac 12). 
%\end {split}
\label {eq:U}
\end {eqnarray}
In (\ref {eq:U}) the matrix elements of single-particle potential $V_{mm'}$
are expressed in terms of complex spherical harmonics $Y_{kq}$, Slater integrals
$F^k$
\begin {equation}
\langle m,m''|V_{ee}|m',m'''\rangle = \sum_k a_k(m,m',m'',m''')F^k, 
\end {equation}
\begin {eqnarray}
%\begin {split}
&a_k(m,m',m'',m''') =  \nonumber\\
&\frac {4\pi } {2k+1}
\sum_{q=-k}^k \langle lm| Y_{kq} | lm' \rangle 
\langle lm'' |Y_{kq}^{*}|
lm''' \rangle, 
\end {eqnarray}
and occupancy $n$, $n$=$\sum_{\sigma m} n^\sigma_{mm}$. For $d$-elements
$k$=0, 2, 4, screened Coulomb parameter $U$=$F^0$, Stoner exchange parameter
$J$=$(F^2+F^4)/14$, with ratio $F^4/F^2$=$0.625$. In the case of strongly
correlated impurity, 
$\Delta V$ in equation (\ref{eq:G_imp}) contains additional contribution according 
to (\ref{eq:U}).

The first step of the proposed CPA method is the construction of the Hamiltonian of the 
system without impurity $\hat H_0(k)$. For that purpose the conventional band structure 
calculation of the pure system was carried out within TB-LMTO-ASA (Tight Binding-Linearized 
Muffin-Tin Orbitals-Atomic Sphere Approximation) code \cite{tb-lmto}.
Then the Hamiltonian of the pure system have been derived in the basis of Wannier functions using 
the projection procedure \cite{Hproj}. 

The second step is the calculation of the difference 
in potentials between the ``impurity'' and ``host'' atoms, $\Delta V$. 
Generally one should simulate the single impurity embedded in the host matrix and calculate the difference of their potentials.
The $\Delta V$ is obtained from the self-consistent band structure calculation 
of the supercell in which one of the atoms in the matrix is replaced by the atom of the impurity.
The supercell should be large enough to have an atom of the matrix which does not have the impurity 
atom in its at least nearest neighbor surrounding. Then the Hamiltonian of the supercell is transformed 
in the Wannier function basis and difference in potentials between impurity atom and the 
outermost of it host atom is derived. We would like to stress here that the supercell approach is used only for  
$\Delta V$ estimation.

In the presented CPA scheme the impurity is supposed to be uniformly distributed among all sites of the matrix.
This implies that the symmetry of the lattice dose not change. The impact of the impurity on the matrix 
is characterized by the value of $\Delta V$. The larger concentration of the impurity $x$ the stronger these 
impact. Note that the concentration $x$ may have arbitrary value.        

\section {Details of the calculation}
It is known from the literature that the NiO-ZnO solid solutions were synthesized in powder 
form at high pressure of 7.7~GPa and at temperature 1470~K in
the crystal structure of rock salt (NaCl)~\cite {synthesis}. Concentration range of single-phase
Ni$_{1-x}$Zn$_x$O solutions is $x=0.3-0.8$. The lattice parameter of the
solution depends linearly on the concentration and varies in the range 4.176~\AA
\ (NiO) -- 4.280~\AA \ (ZnO)~\cite {synthesis}. The neutron diffraction experiments 
revealed that the antiferromagnetic structure AF(II), the same as for NiO, 
preserves in Ni$_{1-x}$Zn$_x$O in  $x=0-0.3$ concentration range~\cite{magstr}.

We present here the results of the electronic structure calculation of strongly
correlated solid solutions Ni$_{1-x}$Zn$_x$O ($x=0-1$) within the coherent
potential approximation, assuming conservation of AF(II) antiferromagnetic structure, 
taking into account the concentration dependence of the lattice parameter in the NaCl structure.
We use a version of the coherent potential approximation introduced in the previous section.

\begin {figure}
\centerline { \includegraphics [angle=270,width=0.8\linewidth]{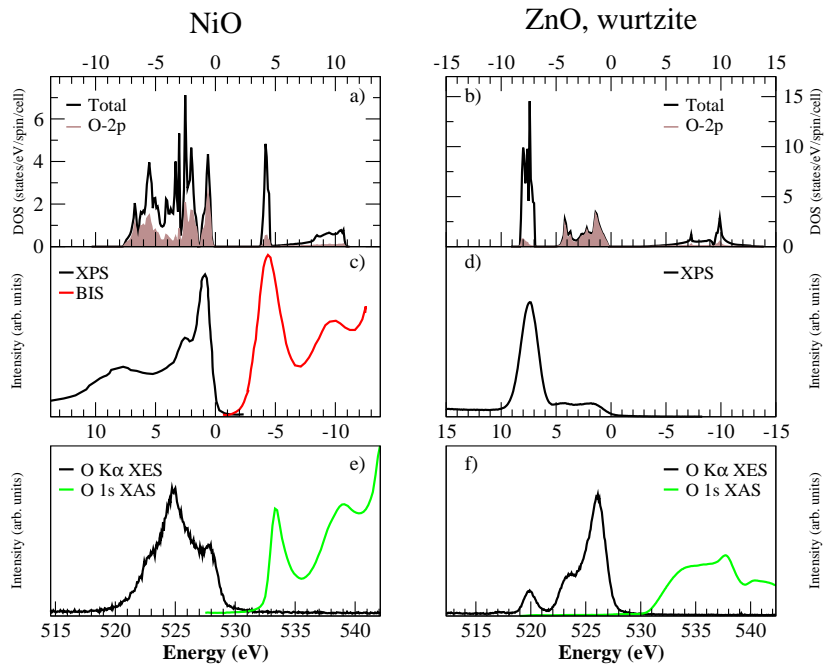} }
\caption {Total and partial O-2p density of states of antiferromagnetic NiO and wurtzite ZnO calculated in
Wannier functions basis in the LDA+U approximation (panels a) and b)) in comparison with the experimental 
XPS VB~\cite{xps_NiO} and BIS~\cite{bis_NiO} (panel c)), O K$\alpha$ XES and O 1s XAS~\cite{nio_spectrum} (panel e)) 
spectra of NiO and  XPS VB~\cite{xps_ZnO} (panel d)), O K$\alpha$ XES and O 1s XAS~~\cite{zno_spectrum} (panel f)) spectra of wurtzite ZnO.
Zero of abscissa for DOSes coincides with the valence band top (panels a) and b)). On the panels c) and d) the abscissa corresponds to the binding energy. 
For panels e) and f) the photon energies are shown on the X axis.} 
\label {fig:ZnO_NiO}
\end {figure}

Different theoretical approaches have been applied to resolve 
correctly electronic structure of pure NiO: LDA+U~\cite{U_J}, generalized gradient approximation 
GGA~\cite{GGA_NiO}, LDA+DMFT~\cite{DMFT_NiO}. The detailed comparison of various theoretical results with 
experimental XES (X-ray Emission Spectroscopy) and XAS (X-ray Absorption Spectroscopy) spectra presented in~\cite{nio_spectrum}. In the current work 
the parameters of Coulomb and exchange interactions for Ni ions in NiO were
taken conventional, $U$=8~eV, $J$=1~eV \cite{U_J}. Obtained within LDA+U (Local Density Approximation+U correction) 
total and partial densities of states for NiO are in good agreement with the previous 
results \cite{U_J} and experimental XES and XAS~\cite{nio_spectrum}, XPS VB (X-ray Photoemission Spectroscopy of Valence Band)~\cite{xps_NiO}  
and BIS ( Bremsstrahlung Isohromat Spectroscopy)~\cite{bis_NiO} spectra (figure~\ref{fig:ZnO_NiO}). 
The calculated within LDA+U characteristics such as
the band gap $\Delta_{calc}$=3.8 eV and the spin magnetic moment $\mu_{calc}$=1.75 $\mu_B$ per Ni ion are in correspondence with
experimental data: $\Delta_{exp1}$=4.3 eV~\cite{exp1}, $\Delta_{exp2}$=3.8 eV~\cite{exp2}, $\mu_{exp1}$=1.64 $\mu_B$~\cite{exp3}, 
$\mu_{exp2}$=1.77 $\mu_B$~\cite{exp4}, $\mu_{exp3}$=1.9 $\mu_B$~\cite{exp5}.

At ambient conditions the most stable structure of ZnO 
is hexagonal wurtzite and it can be transformed to the rock salt structure at pressures 
about 10 GPa~\cite{zno_rev}. The majority of the experimental data concerns 
wurtzite ZnO. The basic features of the XAS and XES~\cite{zno_spectrum} as well as XPS VB~\cite{xps_ZnO} spectra of wurtzite ZnO are reproduced 
in calculation with $U$=7~eV and $J$=1~eV (see figure~\ref{fig:ZnO_NiO}). 
On the contrary to the NiO, in ZnO the valence band near the Fermi level is 
dominated by O-2p states as seen from the shape of DOS O-2p in figure~\ref{fig:ZnO_NiO}. 
The different methods were implicated to overcome a well-known shortcomings of LDA 
for these compound. The comprehensive review could be found in \cite{zno_rev}. 
The obtained in LDA+U calculation value of the band gap is $\Delta_{calc}$=3.0 eV in reasonable agreement 
with experimental estimations $\Delta_{exp}$=3.4 eV~\cite{zno_spectrum}.
Since there are lack of experimental spectra for rock salt ZnO, in what follows 
for ZnO in the NaCl structure we will use the same set of parameters 
(the values of Coulomb and exchange interactions) as for wurtzite. 

There are two formula units in the unit cell of NiO and ZnO with AF(II) magnetic structure. At each concentration 
the LDA Hamiltonian with the corresponding lattice constant~\cite{synthesis} was calculated self-consistently. 
At this stage, the LMTO basis included $s$, $p$, $d$-states 
of Ni(Zn), O and empty spheres, as required by the LMTO code \cite {tb-lmto}. Then the full Hamiltonian 
was projected into the basis of Wannier functions for 3$d$ and 4$s$ states of transition metal ions and 2$p$ states 
of oxygen using procedure described in~\cite {Hproj}. 

To determine the parameters $\Delta$V, calculations of the supercell containing 16
formula units were done. Since the physical idea of
$\Delta$V calculation based on the consideration of a single defect (vanishingly
small impurity concentrations), and we deal with the full range of $x=0-1$ in
Ni$_{1-x}$Zn$_x$O, a study was carried out from two directions: when Zn is an
impurity in NiO, $x=0-0.5$, and when there is an impurity of Ni in ZnO, $x=0.5-1$. 

In a self-consistency coherent potential loop the parameter of inverse
temperature was chosen to be $\beta$=40~eV$^{-1}$, which corresponds to the room
temperature, the Matsubara frequency cutoff was set to E$_{cut}$=1500~eV.

\section {Results and discussion}
\begin {figure}[!]
\centerline { \includegraphics [clip=false,width=0.8\linewidth]{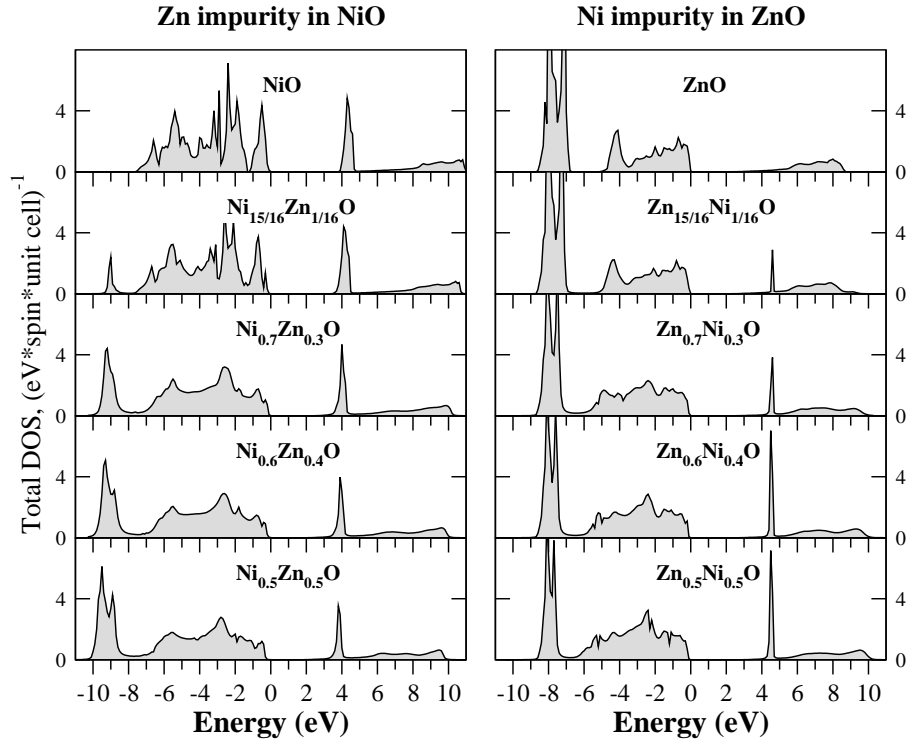} }
\caption {Energy spectra of antiferromagnetic Ni$_{1-x}$Zn$_x$O solid solutions calculated 
in the coherent potential approximation. Zeros of abscissa coincide with the top of the valence band.}
\label {fig:es}
\end {figure}

The calculated results for the total DOS of solid solutions
Ni$_{1-x}$Zn$_x$O are shown in figure~\ref {fig:es}, where the results 
for Zn impurity in NiO, $x=0-0.5$, are displayed in the left column 
and for Ni impurity in ZnO, $x=0.5-1$, in the right one. 
In the case of undoped NiO (left top panel in figure~\ref {fig:es}) the valence and conduction bands are formed by hybridized
O $2p$- and Ni $3d$-states. The width of the valence band is $\sim7.5$~eV. Under Zn
doping emerges an additional band centered at the energy $\sim-9$~eV and formed by impurity
Zn $3d$-states. An increase of Zn concentration leads mainly to
narrowing of the valence band of undoped NiO and to growth of Zn $3d$-peak.
Simultaneously, the width and the height of conduction band peak are decreased
due to reduction of the number of unoccupied $3d$-electrons originated from Ni.

Valence band of pure rock salt ZnO (right top panel in figure~\ref {fig:es}) consists of
two parts: Zn $3d$-band of $\sim2$~eV width centered at energy  $\sim-8$~eV and
O $2p$-band with admixture of Zn $3d$-states extended in ($-5\div0$)~eV
interval. The increase of Ni impurity concentration leads to appearance and
growth of unoccupied conduction band peak, broadening of O $2p$ valence band,
and decrease of Zn $3d$-band.

Experimental data of the optical absorption edge \cite {expgapZnO} show that 
rock salt ZnO is an indirect semiconductor with a band gap of 2.45$\pm$0.15~eV at 13.5~GPa. 
An intense direct transition occurs at higher energy $\sim$4.5~eV at 10~GPa. 
The energy gap obtained within LDA+U calculation for pure ZnO (right upper panel of 
figure~\ref{fig:es}) is about 2.2 eV in agreement with experimental results.

From figure~\ref{fig:es}(left panel) one can see that the energy gap in Ni$_{1-x}$Zn$_x$O 
solid solutions with $x=0-0.5$ undergoes only slight reduction of about 0.25 eV going from pure NiO to 
Ni$_{0.5}$Zn$_{0.5}$O. The photoluminescence and photoluminescence excitation of NiO and Ni$_{1-x}$Zn$_x$O 
solid solutions with NaCl crystal structure were investigated in~\cite{Sokolov12}. On the basis of  
photoluminescence excitation spectra for this solid solutions one can conclude that the value of 
the band gap in the series NiO-Ni$_{0.5}$Zn$_{0.5}$O-Ni$_{0.3}$Zn$_{0.7}$O displays minor changes not more than 0.2 eV~\cite{Sokolov13}.
This experimental finding is in good agreement with the present CPA results.  
\begin {figure}[!]
\centerline { \includegraphics [width=0.5\linewidth]{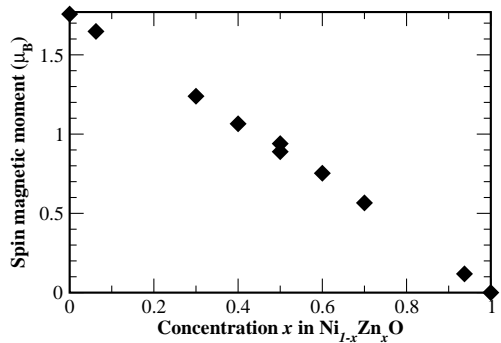} }
\caption {Spin magnetic moment of effective $d$-ion in Ni$_{1-x}$Zn$_x$O solid solutions 
calculated in the coherent potential approximation.}
\label {fig:mm}
\end {figure}
The values of spin magnetic moment of effective $d$-ion in Zn impurity in NiO and Ni impurity in ZnO are shown in figure~\ref{fig:mm}. 
The small difference of calculated results 0.05~$\mu_B$
at "meeting" concentration $x=0.5$ validates the coherent potential approximation scheme used in this work.

Magnetic moment values presented in figure~\ref{fig:mm} are calculated basing on the single-site Green function (\ref{eq:G_iw}) 
for the effective $d$-shell. It's values decay linearly from 1.77~$\mu_B$ in undoped NiO to $0$ in undoped ZnO. Meanwhile 
the magnetic moments for the host $d$-Ni determined from the host Green function (\ref{eq:G_host_U}) are practically the same 
in the concentration range 0$<x<$0.5 (the difference between them is less than 0.02 $\mu_B$). This agrees well with 
the neutron diffraction data indicating no change in the magnetic moment despite of the magnetic ion dilution (see figure 5 in~\cite{magstr}). 

\section {Conclusions}
The method for electronic structure calculation of disordered strongly correlated
systems is developed on the basis of the coherent potential approximation.
It is applied for investigation of the Ni$_{1-x}$Zn$_x$O solid solutions. 
The novelty of this study lies in the fact that the CPA is applied to the disordered strongly correlated 
transition metal oxide system for the first time. The energy spectra and spin magnetic moment values 
have been calculated assuming NaCl crystal structure and antiferromagnetic order of $d$-ions in the whole concentration range
$x$=$0-1$ in Ni$_{1-x}$Zn$_x$O. Strong correlations of $d$-ions were treated in the static mean-field limit. 
The values of the band gap in the Ni$_{1-x}$Zn$_x$O solid solutions exhibit insignificant changes in agreement with results 
of the photoluminescence excitation. The decrease of the spin magnetic moment value of the effective $d$-ion 
is found to be linear with the increase of concentration $x$.

\section *{Acknowledgements}

The authors thank Yuri A. Babanov and Victor I. Sokolov for useful discussions.
The study was partially supported by Program of UB RAS, project No 12-P-2-1021 (MAK), 12-I-2-2040 (EZK) and 
RFFI 13-02-00374, MK 3443.2013.2 (ZVP). Part of the calculations were performed on the “Uran” cluster of the
IMM UB RAS.

\end {document}